\documentclass[conference]{IEEEtran}
\IEEEoverridecommandlockouts
\usepackage{cite}
\usepackage{amsmath,amssymb,amsfonts}
\usepackage{algorithmic}
\usepackage{graphicx}
\usepackage{textcomp}
\usepackage{xcolor}

\usepackage{cite}
\usepackage{amsmath,amssymb,amsfonts}
\usepackage{algorithmic}
\usepackage{graphicx}
\usepackage{textcomp}
\usepackage{xcolor}
\usepackage{booktabs}
\usepackage{lipsum}
\usepackage{subcaption}
\usepackage{fancyhdr}
\usepackage{inconsolata}

\def\BibTeX{{\rm B\kern-.05em{\sc i\kern-.025em b}\kern-.08em
    T\kern-.1667em\lower.7ex\hbox{E}\kern-.125emX}}

\begin{document}


\title{Robust web element identification for evolving applications by considering visual overlaps}

\author{\IEEEauthorblockN{Michel Nass, Emil Alégroth, Robert Feldt}
\IEEEauthorblockA{\textit{SERL} \\
\textit{Blekinge Institute of Technlogy}\\
Karlskrona, Sweden}
\and
\IEEEauthorblockN{Riccardo Coppola}
\IEEEauthorblockA{\textit{DAUIN}\\
\textit{Politecnico di Torino}\\
Turin, Italy}
}

\maketitle

\begin{abstract}

Fragile (i.e., non-robust) test execution is a common challenge for automated GUI-based testing of web applications as they evolve.
Despite recent progress, there is still room for improvement since test execution failures caused by technical limitations result in unnecessary maintenance costs that limit its effectiveness and efficiency.
One of the most reported technical challenges for web-based tests concerns how to reliably locate a web element used by a test script.

This paper proposes the novel concept of Visually Overlapping Nodes (VON) that reduces fragility by utilizing the phenomenon that visual web elements (observed by the user) are constructed from multiple web-elements in the Document Object Model (DOM) that overlaps visually.

We demonstrate the approach in a tool, VON Similo, which extends the state-of-the-art multi-locator approach (Similo) that is also used as the baseline for an experiment. 
In the experiment, a ground truth set of 1163 manually collected web element pairs, from different releases of the 40 most popular web applications on the internet, are used to compare the approaches' precision, recall, and accuracy.

Our results show that VON Similo provides 94.7\% accuracy in identifying a web element in a new release of the same SUT.
In comparison, Similo provides 83.8\% accuracy.

These results demonstrate the applicability of the visually overlapping nodes concept/tool for web element localization in evolving web applications and contribute a novel way of thinking about web element localization in future research on GUI-based testing.


\end{abstract}


\section{Introduction} \label{intro}
In modern software engineering, test automation is a key activity, where automated tests are used to continuously monitor the software's quality and provide frequent feedback to developers~\cite{mahmud2014design}.
However, much of this automation has been restricted to lower-level testing such as unit and integration tests~\cite{olan2003unit}.
Higher level testing, particularly with graphical user interface (GUI) tests, are still mostly manual and therefore a costly activity in practice~\cite{grechanik2009maintaining, grechanik2009creating}.

While GUI tests can be used to verify the correctness of the GUI's appearance, the focus of many GUI tests is on verifying functional correctness of the system under test (SUT)~\cite{alegroth2018continuous}, i.e. system testing through the SUT's GUI~\cite{berner2005observations}.
However, despite continued research since the 1980s, several key challenges remain and limit the widespread adoption of automated GUI testing~\cite{nass2021many}.
One of these challenges is the robust identification of GUI elements. 
This issue has been described for many domains, and several approaches have been proposed to increase the robustness of GUI element localization~\cite{leotta2014reducing, leotta2016robula+, yandrapally2014robust, montoto2011automated}. 
Robustness is, in this instance, defined as the correct identification of a web element when it is available and reporting no match when a web element is unavailable. 
This property is particularly important as automated web application tests are typically used for regression testing as software systems, and their GUI elements, change and evolve~\cite{engstrom2010qualitative}.

Despite its importance, research has had marginal success in solving the challenge of robust GUI element localization~\cite{nass2021many}. 
Instead, much research has been focused on extending GUI testing technologies, only utilizing the already available web element localization solutions---example of such extensions are test generation and GUI ripping~\cite{vos2015testar,nguyen2014guitar}. 
Some research, however, has been made investigating new types of locators, e.g., image recognition~\cite{yeh2009sikuli}, or multi-locators~\cite{leotta2016robula+}.
However, we still consider web element localization an unsolved challenge~\cite{nass2021many}, warranting more research to improve the general robustness and maintainability of available GUI testing techniques and tools.

Nass et al. proposed an approach called similarity-based web element localization (Similo) \cite{nass2022similarity}, which calculates a weighted similarity score between the target web element in a previous version, and all web elements (candidates), in a revised version, of a web application.
The target web element contains the desired properties (e.g., attributes) that are compared with each candidate.
They compared the Similo approach with the multi-locator approach proposed by Leotta et al. \cite{leotta2015using} and found that Similo can correctly locate more target web elements than the multi-locator when evaluating web elements extracted from 40 commonly used homepages.


This study presents the novel concept of Visually Overlapping Nodes (VON).
The concept makes use of the structure of how modern web applications are constructed (i.e., as hierarchies of components with specific attributes and characteristics) and how this structure is formalized in the document object model (DOM)~\cite{wood1998document}.
Our investigations show that multiple DOM nodes often point to the same visual element in the rendered GUI, implying that any of these nodes, if found to be a match to the sought visual element, will constitute a valid match if used for a graphic validation of the SUT.
DOM nodes (typically arranged in a hierarchy) that represent the same visual element (e.g., a button or menu item) are, in this paper, referred to as visually overlapping nodes (VON).
As an example, a menu item that is represented by an anchor tag ($<$a$>$) containing two span tags ($<$span$>$) in the DOM.
All three DOM nodes visually appear as the same element (as observed by the user) even though they are three separate nodes.
The benefit of this approach is that it increases the chance to find at least one good match for a provided target, thus leading to more robust web element identification.

\begin{figure}[t]
\includegraphics[width=\columnwidth]{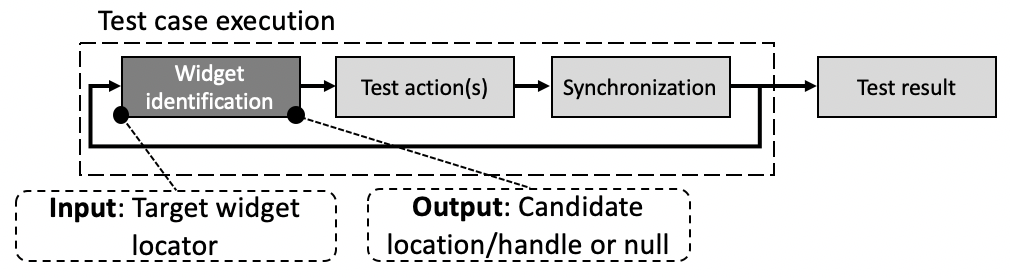}
\caption{Graphical representation of the GUI test case execution process, highlighting the step (web element identification) that is studied in this work.}
\label{test_process}
\end{figure}

While many other essential activities need to be in place for robust GUI-based test automation, we here focus on the vital aspect of GUI web element identification.
If the identification is not robust, later steps of the chain cannot compensate for this.
This focus is visualized in Figure \ref{test_process}.

As such, the main contributions of this work are:
\begin{itemize}
    \item Insights into the relative power of different web element attributes for web element localization;
    \item A generally applicable, yet novel, concept called Visually Overlapping Nodes (VON);
    \item An improved version of similarity-based web element localization (Similo) that implements VON (VON Similo).
\end{itemize}

The continuation of this paper is structured as follows.
First, in Section \ref{background}, we present related work and give more detailed descriptions of the technologies evaluated in this study.
Section \ref{methodology} then explains the study's experimental setup.
We present the results in Section \ref{results}, which describes the findings of our experiment.
These findings are discussed in Section \ref{disc} before the paper is concluded in Section \ref{conc}.


\section{Background and Related Work} \label{background}


In the GUI-based web application testing discipline, a web-element locator is defined as a method, function, approach or algorithm that can locate a web element in a given web page according to a specific parameter. State-of-the-art techniques and tools typically make use of conditions on the attributes of the web elements in the HTML DOM tree (e.g., ids, attributes, or class names).
Popular testing tools (e.g., Selenium) provide the possibility to use XPath or CSS expressions to locate elements on the web page \cite{bruns2009web}.
We refer to these types of locators as \emph{single-locators}.

Single-locators are a reported source of \emph{fragility} for test suites.
Fragility is defined as the lack of robustness of the locators to changes in the GUI definition of the SUT.
Test script fragility typically manifests through test cases that fail not because of functional misbehaviors in the SUT but because of the inability of the testing engine to find the web elements needed during the test sequences using existing locators \cite{coppola2018mobile}.
When a locator cannot be found in the DOM tree of the current web page, it is typically referred to as a \emph{Broken locator} \cite{leotta2015using}.

Given the increased complexity and high pace of change of modern web applications, it is highly likely that the attributes or the XPaths of web elements are modified between two versions of the same web application.
If single-locators are used, any such change would result in failed localization attempts and require additional effort by the testers to repair the broken test suites.

Leotta et al. proposed the multi-locator approach \cite{leotta2015using} to limit the fragility of single-locators in web application testing.
The approach evaluates multiple locators and uses a voting procedure between the single-locators to improve the accuracy of locating the correct web element in a web-page, thereby improving the robustness and reducing the script maintenance cost.


Similo is another approach of multi-locators based on a weighted similarity score computed on the differences between the locator parameters of the web element to locate and each of the web elements on the current web page \cite{nass2022similarity}.
Unlike the multi-locator approach proposed by Leotta et al. \cite{leotta2015using}, which uses five single-locators that uniquely identify precisely one (or none) web element each, Similo combine comparisons on multiple attributes in a single score. Such score can then be used to rank the possible candidate widgets for the current target, thereby finding multiple possible alternatives instead of a single result as Leotta's multilocator. The algorithm can then select as valid matches all the candidates for which the score is above a given threshold, or select the highest-rated candidate as the single match for the target.

Similo can take advantage of locators that pinpoint more than one web element.
For example, Similo can use an absolute XPath that pinpoints exactly one web element in the DOM, but unlike the multi-locator proposed by Leotta et al., Similo can also use the CSS selector ".name" to find all web elements that contain a specific class name even if the query results in several matching web elements.

\begin{figure}[t]
\includegraphics[width=\columnwidth]{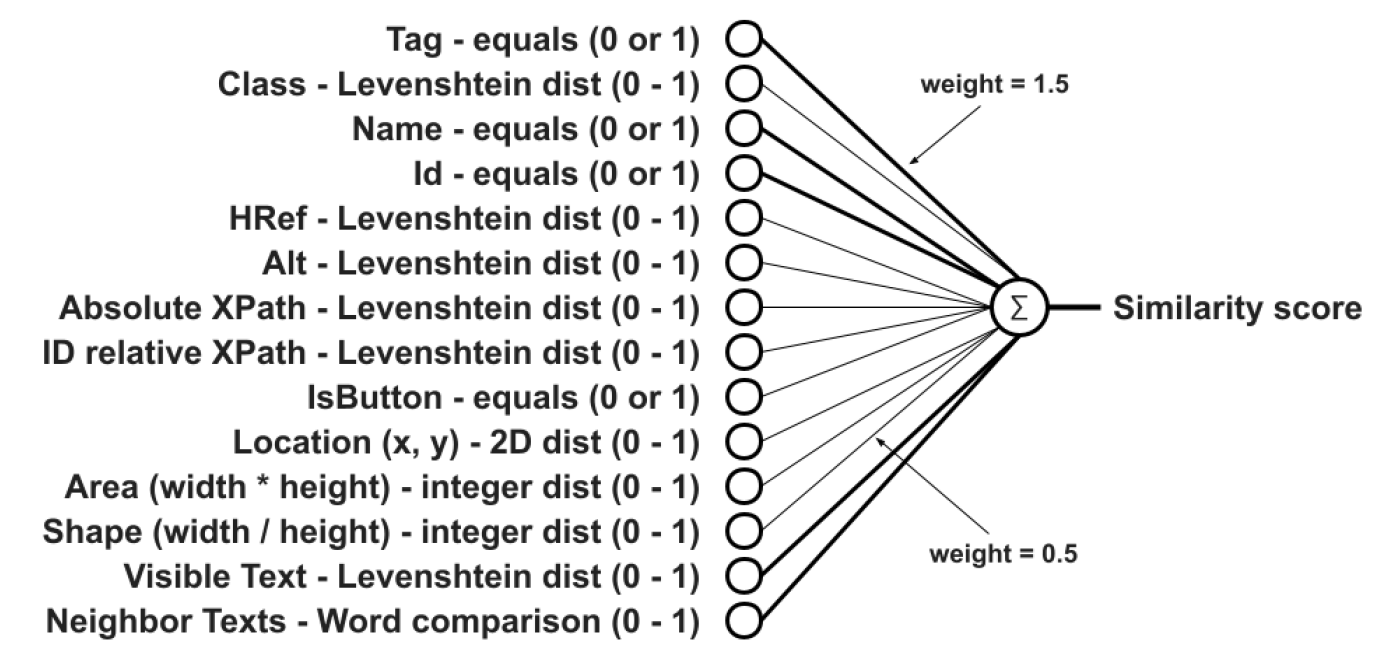}
\caption{Graphical representation of the computation of similarity score between two different sets of locator parameters.}
\label{similo_original}
\end{figure}

In the Similo study \cite{nass2022similarity}, currently available as a preprint at arXiv, the authors used 14 different locator parameters with corresponding comparison operators and weights summarized in Figure \ref{similo_original}.
The locator parameters Tag, Class, Name, Id, HRef, Alt, XPath and ID relative XPath, Location, Visible Text are collected directly from the DOM-tree.
IsButton is a derivate boolean parameter, that was set to true or false according to the values of the attributes Tag, Type, and Class.
Neighbor Texts contain a space-separated text of words collected from the visible text of nearby web elements.
Specific comparison operators that return a value between zero and one (or exactly zero or one) were selected for each locator parameter.
Some locator parameters were compared by the Java method equalsIgnoreCase (e.g., Tag, Name, and Id).
Others were compared using Levenshtein distance, word comparison, or Euclidean distance.
Detailed information about the comparison operators can be found in the Similo paper \cite{nass2022similarity}.

The weights for each locator parameter (and comparison operator) were assigned based on their respective stability and uniqueness found by the COLOR study by Kirinuki at al. that used a similar selection of locator parameters when evaluating four open-source web applications \cite{Kirinuki2019COLORCL}.
The COLOR approach considers various properties such as attributes, positions, texts, and images to propose a repair unlike Similo that approaches the same problem in a preventive way (i.e., before the script failure occurs).
In Similo, the locator parameters were divided into two groups based on the corresponding weights from the COLOR study.
Locator parameters with a higher stability and uniqueness were assigned the weight value 1.5 (bold lines in Figure \ref{similo_original}) and the remaining were assigned the value 0.5 (thin lines in the graphical representation).

\section{Visually overlapping nodes approach}  \label{von_approach}

In this section, we propose an approach to enhance the current state of the art in multi-locators for web application testing with an improved version of similarity-based web element localization (Similo) that implements visually overlapping nodes (VON Similo).

A characteristic of modern web applications is that they are built from elements that contain other elements, e.g., a text label can be contained in a button that is itself contained in a div tag, which in turn may be placed in another container, such as a frame.
This hierarchical structure is modeled within the DOM, which is used by web browsers and GUI testing tools to render, identify or interact with elements.
From a DOM perspective, each element is considered a unique entity, as it is described by a DOM node identifiable through a unique absolute XPath.
However, from a visual perspective, multiple nodes --- due to size, placement, and content --- represent the same visual element, e.g., a button.
This feature of modern web applications thereby implies the existence of a many-to-one connection between DOM nodes and visual elements.
In this paper, all the DOM nodes that belong to the same visual element are referred to as visually overlapping nodes (VON).

\begin{figure}[t]
\includegraphics[width=\columnwidth]{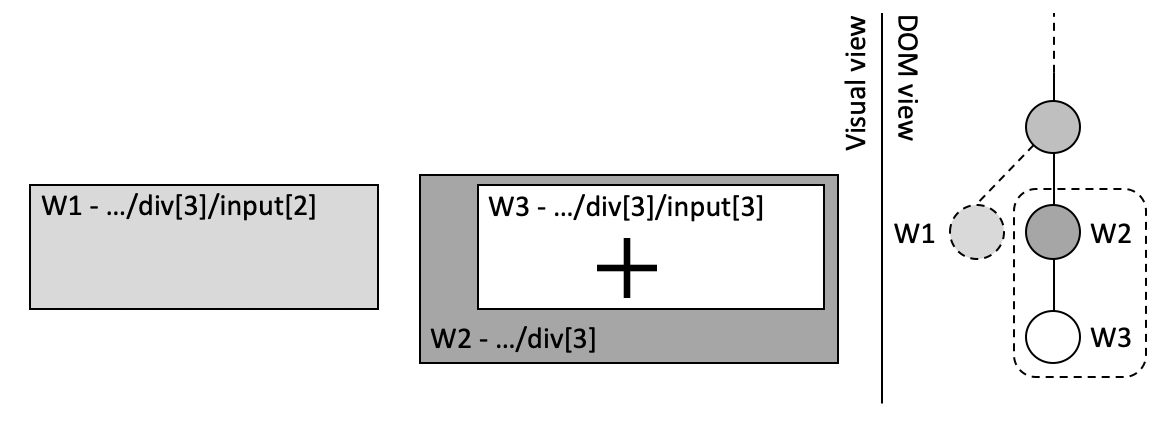}
\caption{A visualization of a hierarchy of web elements represented both visually and from a DOM perspective. It shows that although W2 and W3 are unique entities, they appear to be the same visual component or, at least, overlap visually.}
\label{eq_visual}
\end{figure}

Fig. \ref{eq_visual} visualizes this many-to-one correspondence, showing how web elements can be contained in other web elements yet, from a visual perspective, occupy the same area of the screen.
In this case, while web elements W2 and W3 are represented with different XPaths in the DOM, both visually point to the same (or nearby) screen area or web element.
This phenomenon implies that both web elements (W2 and W3) are equally correct candidate elements (i.e., elements available on the current web page) for a target element (i.e., the element that contains the properties that we are looking for) pointing to that screen area.
We refer to this phenomenon as visually overlapping nodes (VON), where property-based localization approaches, like Similo, can utilize VON to increase the number of correctly located candidate web elements.
Hence, rather than relying on finding one specific DOM node, or absolute XPath, any located DOM node that belongs to the same visual element is deemed a correct match.
This approach mitigates test execution fragility by, as mentioned, increasing the number of candidate DOM nodes that can be matched when identifying one and the same web element.
Essentially, the approach will be more robust as long as only one or a subset of these nodes change between two revisions.

Formally, we define the approach to identify equivalent web elements as: 

Given a web element W1, we define the set of equivalent web elements e$_0$, … e$_N$ as the set of web elements that satisfy the following properties:

\begin{enumerate}
\item The ratio between the overlapping areas of the web elements on the screen, and the union of the areas of the two web elements, is higher than a set threshold value. Such ratio is computed as, 

\[ \frac{\cap(R_1, R_2)}{\cup(R_1, R_2)} \]

where: $R_1$ and $R_2$ are the rectangles occupied on the screen by the two web elements; the set intersection symbol indicates the size (in pixels) of the common area occupied on screen by $R_1$ and $R_2$, and the set union symbol indicates the size (in pixels) of the union of $R_1$ and $R_2$.

By experimenting with different threshold values to identify visually overlapping nodes, we finally selected 0.85 as a value for the threshold that allowed us to avoid a definition of visual overlap that was too loose (thereby considering visually separate GUI elements as overlapping) or too strict (thereby finding none or a few visually overlapping nodes).

\item	The center of the web element W1 is contained in the rectangle R1.
\end{enumerate} 

The addition of the VON concept transforms the locator parameters (e.g., tag. id, or xpath) that are stored and utilized by the Similo algorithm. 
Each locator parameter value is no longer a single value but is instead substituted by a list of equivalent values collected from all the visually overlapping web elements of a DOM node.

The VON Similo score substitutes the comparison functions of the original Similo, with the following function:

Given the set ew$_1$ (set of web elements equivalent to w$_1$: e$_{1.1}$, e$_{1.2}$, ... e$_{1.N}$) and ew$_2$ (set of web elements equivalent to w$_2$: e$_{2.1}$, e$_{2.2}$, ... e$_{2:M}$), and the values for a specific attribute of the equivalent web elements, VON Similo computes the similarity between all the possible combinations of visually overlapping web elements for a specific attribute.
The maximum of these similarity values is the VON Similo similarity score for that attribute in the comparison, representing the best possible match to the target web element.
However, while the match may not be the target web element, this approach ensures that the coordinates of the match overlap with the intended target.
As such, from a visual perspective, the target element is identified.

\begin{figure}[t]
\includegraphics[width=\columnwidth]{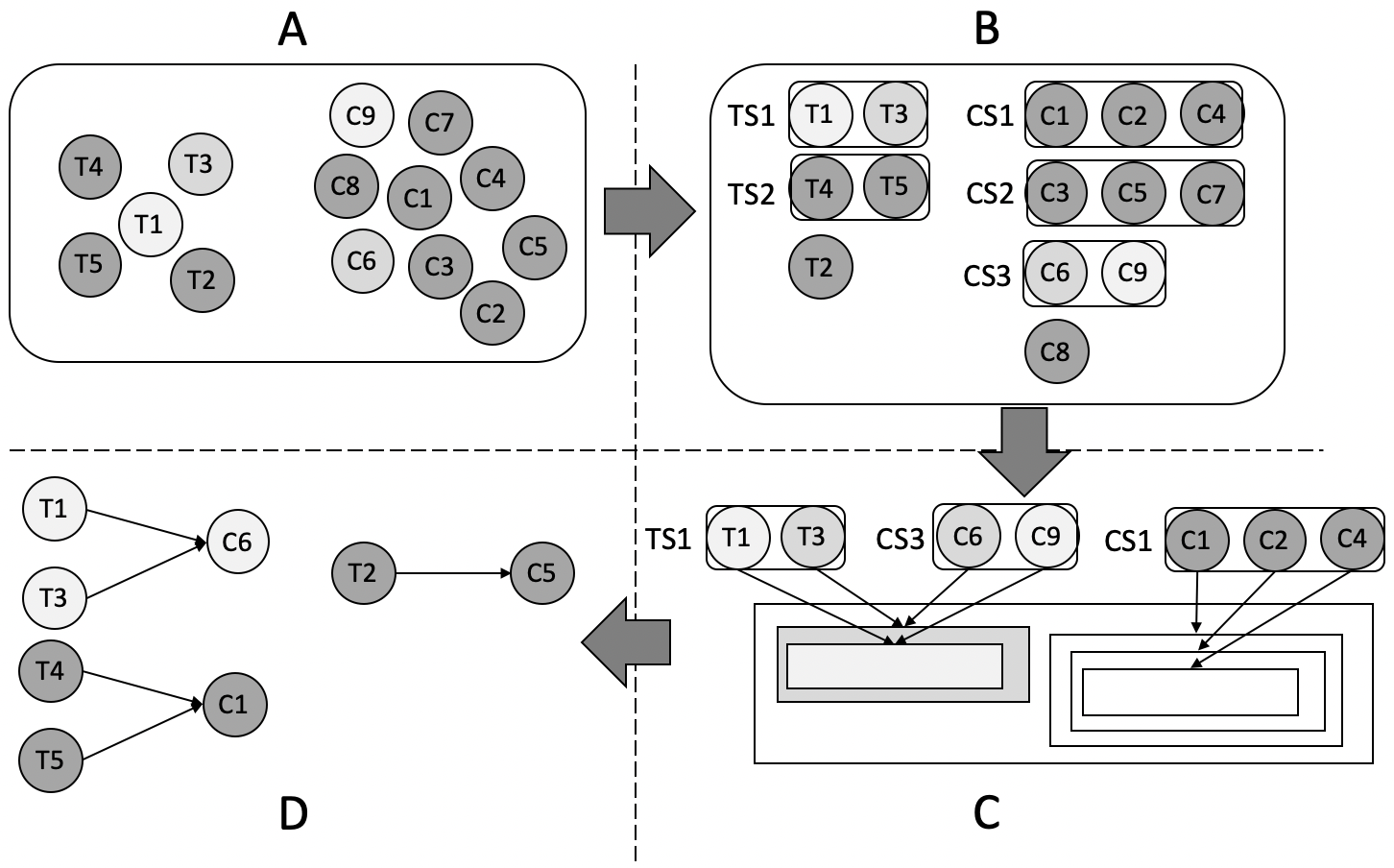}
\caption{Visualization of how visually overlapping nodes are implemented in VON Similo.}
\label{similo_eml}
\end{figure}

Figure \ref{similo_eml} presents a visualization of the process of how visually overlapping nodes are utilized in VON Similo.
In the first step, denoted A in the figure, a set of target web elements (denoted T$_x$$\in$TS) and candidate web elements (denoted C$_y$$\in$CS) are available, where $|$TS$|$ $\leq$ $|$CS$|$.
In the second step, denoted B in the figure, a pre-analysis of TS and CS is performed, clustering all target and candidate web elements according to the visual web elements they are associated with, using the formula presented previously in this section.
The outcome of the pre-analysis are clusters TS$_1$-TS$_i$ and CS$_1$-CS$_j$ containing components with overlapping target areas on the screen but otherwise with an unknown overlap in terms of locator properties.
In step 3, denoted C in the figure, each target web element T$_x$$\in$ $TS_i$ is compared to every other candidate web element C$_y$$\in$ $CS_j$, and a similarity score is calculated.
After the comparison, the maximum similarity score of each cluster is kept and associated with all target web elements T$_n$-T$_m$$\in$TS$_i$, resulting in a mapping between T$_i$ and C$_j$ as visualized in the last step of the figure, denoted D.
This mapping implies that (from a DOM perspective) a given target web element T$_i$ may not be mapped to the candidate component C$_i$=T$_i$ that was initially used when the target was set in the previous version of the SUT.
In Figure \ref{similo_eml}, T$_1$ is equivalent to C$_9$ but is mapped to its parent component C$_6$.
However, from a visual perspective, this is irrelevant since both C$_6$ and C$_9$ point to a visual area that is overlapping with the one containing T$_1$.

The benefits of this approach is that the number of valid matching candidates increases, implying that for clusters where a target web element can not be mapped to a suitable candidate, a candidate can still be associated.
This increases test robustness by reducing the number of failed test actions caused by minor changes to components that would otherwise be considered false positive test results.
One could argue that this approach would increase the number of false positives, or false negatives, by mapping targets to incorrect candidates.
However, as we will show, no such trends were identified.

\section{Empirical Evaluation} 
\label{methodology}

In this section, we report the goal, the research questions, the methodology, and the results of the empirical evaluation we performed when comparing VON Similo to the baseline.
Similo was selected as the baseline since a previous study \cite{nass2022similarity} (currently only available as a preprint) showed that it was able to locate more web elements when compared to the multi-locator approach proposed by Leotta et al. \cite{leotta2015using}.

\subsection{Goal}

The study's high-level goal is to evaluate the efficiency of VON-based web element locators when applied to web applications. The results are interpreted from the perspective of software testing procedures needing methods to automatically locate web elements in evolving GUIs with high levels of accuracy.

To perform a comparison with the state-of-the-art, we performed an analysis of the accuracy and effectiveness for the base Similo algorithm, that was not performed in the original paper. We also complement the original study with an analysis of the aptness of DOM attributes for widget localization.


\subsection{Research Questions and Metrics}

\begin{itemize}

\item \textbf{RQ1}: Which are the most frequently used element attributes in web applications?
\item \textbf{RQ2}: What is the effectiveness of similarity-based web element localization (Similo)?
\item \textbf{RQ3}: What is the effectiveness of VON-based web element localization (VON Similo)?
\item \textbf{RQ4}: Which type of multi-locator (Similo or VON Similo) performs better in terms of accuracy?
\end{itemize}

The objective of RQ1 is to corroborate the selection of an optimal set of attributes for the multi-locator approach employed in Similo (attributes shown in Figure \ref{similo_original}).
We chose these attributes based on related literature.
Still, it is unknown how often the attributes are populated with data on a generic web page and, thereby, what value they give to the employed calculation of similarity.
It is further unknown if any attribute, which could provide value, is not part of this set.
To answer RQ1, we compute two different metrics: the \emph{Relative Non-null} values for each attribute and the \emph{Variability} of each attribute.

The objective of RQ2 is to extend and complement the original study performed on Similo \cite{nass2022similarity}.
Our purpose is to compute Accuracy, Precision, and Recall for the original similarity-based web element localization approach, to serve as a baseline for our further evaluations.
Real-world web applications are used, where we can observe minor and significant changes to the site's appearance and functionality. 
The standard Accuracy, Precision, and Recall measures are calculated based on the use of a human oracle by mapping target web elements from the old version of the web application to the new web application.

The objective of RQ3 is to evaluate the accuracy with which VON Similo can locate the correct web element (from a visual perspective ), given a target web element, on a new release of the same web application.
To answer RQ3, we resort again to measuring three standard measures, Accuracy, Precision, and Recall. 

Finally, the objective of RQ4, as a side-objective of the measurement of the Accuracy, Precision, and Recall provided by both types of locators, is to provide a comparison between them.

\subsection{Methodology} 
This section presents the steps we took to evaluate the research questions in a controlled experiment.

A replication package is available as an open-source repository \footnote{https://figshare.com/s/e63c3679e925730397ac}.

We will present threats to the validity of the study design in Section \ref{threats}.

\begin{enumerate}
\item \emph{Application Collection}: To gather a set of experimental subjects for our evaluations, we collected pairs of versions of the same webpage inside the same web application.
The objective was to emulate the maintenance of web application by using target web elements from an old site version and identifying them on a new version.

To avoid biased selections, we selected the 40 top-rated web application in the United States from the Alexa ranking web application\footnote{http://www.alexa.com}\footnote{Since the experiment, the web application (i.e., alexa.com) has been discontinued and is no longer publicly available.}.
Since the list contained one case of mirrored sites (different URLs leading to the same web application), we used only one of them.
We also excluded one web application with adult content due to ethical reasons. The final list of 38 web applications can be found in the replication package.

\item \emph{Application Version Selection}: The Internet Archive web application was used\footnote{http://web.archive.org}, to acquire the old version of the web application chosen in the previous step.
To further reduce potential biases, we choose to replicate a design employed by Leotta et al., selecting as the newer version (\emph{v2}) of each web application a version published in December 2020 and as the older version (\emph{v1}) a version dated R months backward in time (with R randomly varying between 12 and 60 months, 36 months on average). 
This choice ensured enough time had elapsed between releases to see graphical and functional differences between the two versions of the web application, enabling us to evaluate the web element finding ability of the approach over time for both minor and significant changes.
We perceive minor changes to have higher construct validity for regular operations in practice.
Still, we do not want to exclude more significant changes, which can occur when companies re-brand or make more extensive technological updates to their web applications.

\item \emph{Application scraping}: We developed a Selenium-based web scraper to analyze the distribution of the most commonly used web elements and attributes and their distribution among web applications.
The scraper collects all attribute values for all web elements for a web page and stores them for further analysis.
This scraping is achieved by a script that cycles all web elements in the DOM tree and extracts values defined in the W3C list of HTML attributes.
The scraped information is stored in CSV files.

Based on the scraped information, we were able to collect the following statistics to answer RQ1: (i) the number of occurrences of non-empty attributes, i.e., different than "" (empty) or \emph{null}; (ii) the variability of the values, i.e., the ratio between the number of different values divided by the total number of non-null, non-empty values.
We recognize that the latter metric has more or less inherent variability for specific attributes, e.g., labels. Still, since we did not look at the content for this evaluation, only the variation in content, we perceive this effect to be minor.

\item \emph{Correspondent web element Selection}: To acquire target web elements and oracles for the automated evaluation, we manually selected web elements from each of the 40 web application homepages.
We only selected elements from the start page since the Internet archive only stores static pages, meaning that javascript, databases, etc., do not always work.
We were only interested in the web element finding ability of the approach and perceived that it is unlikely that web elements on other pages on a web application have a significantly different distribution than those in the home pages. 
The sample of web elements was also analyzed to verify that we had acquired a comprehensive set of different types of web elements.

The web elements that were chosen for the evaluation had to comply with one or several of the following attributes: (i) were possible to perform actions on (e.g., anchors, buttons, menu items, input fields, text fields, check-boxes, and radio-buttons); (ii) can be used for assertions or synchronization (e.g., top-level headlines); (iii) belong to the core functionality of the web application homepage; (iv) are present in both versions of the web application homepage.
The rationale for (iv) is given by the study's objective to look at the approach's ability to find web elements over different versions of the web applications.
Through this manual selection, we devised a set of 442 matches (pairs of corresponding web elements) for the experiment.


\item \emph{Equivalent web elements Selection}: The number of manually-identified matches is then expanded by applying the equivalence definition according to the formula in Section III of the present paper.
By considering the set of equivalents of both web elements in each of the 442 pairs, we came up with an extended set of 1163 matches.

We finally define a balanced test set for applying the Similo and VON Similo algorithms.
To build the test set, we include the aforementioned set of 1163 matches and 1163 randomly-selected non-matching web element pairs.
We take the same number of matching and non-matching web element pairs for each considered web application.

\item \emph{Match Definition}: In the experiment step, for each pair of web elements (either matching or non-matching), we compute the Similo similarity score and the VON Similo score.
The scores are normalized in the range [0, 1]. 

No normalization step was performed in the original formulation of the Similo locators.
The algorithm, instead, computed the similarity scores for all candidate web elements and ranked them with no normalization steps.
For our purposes, it was necessary to add a normalization step to compare the performance of Similo with that of VON-based locators.

A web element of the new web application is considered a match to the target web element if the Similo (or VON Similo) score is higher than a threshold. The threshold is provided as a variable parameter to the algorithm, which is necessary since both algorithms would otherwise always return a web element (even with a meager similarity score).

\item \emph{Analysis}: The results were then analyzed using formal and descriptive statistics to identify the metrics used to answer the study's research questions.
For this study, we use the target web element (W$_t$) to describe the sought web element taken from the older web application version.
In turn, the correct candidate web element (W$_c$) is sought in the new version of the web application such that W$_c$ $\in$ WS, where WS is the set of all web elements in the newer version of the web application.
Given these definitions, we get the following outcomes of a web element localization:

\begin{itemize}
    \item \textbf{True positive (TP)} - When there is a found match such that W$_t$ = W$_c$ when W$_c$ $\in$ WS.
    \item \textbf{False negative (FN)} - When there is no match found such that W$_t$ $\neq$ W$_c$ despite W$_c$ $\in$ WS.
    \item \textbf{True negative (TN)} - When there is no match found such that W$_t$ $\neq$ W$_c$ when W$_c$ $\notin$ WS.
    \item \textbf{False positive (FP)} - When there is a match found such that W$_t$ = W$_c$ despite W$_c$ $\notin$ WS.
\end{itemize}
Counting the occurrences and distributions of the above measurements provides us with the information required to answer all the study's research questions. Based on such definitions we in fact compute the following derived metrics: \emph{Precision}, as TP/(TP+FP); \emph{Recall}, as TP/(TP+FN); \emph{Accuracy}, as (TP+TN)/(TP+FP+TN+FN).

\end{enumerate}

\section{Results}

In this section we report the results measured to answer the research questions defined for the study. 

\subsection{RQ1 - Ideal set of attributes for VON Similo} \label{results}

\begin{figure}
    \centering
    \includegraphics[width=\columnwidth]{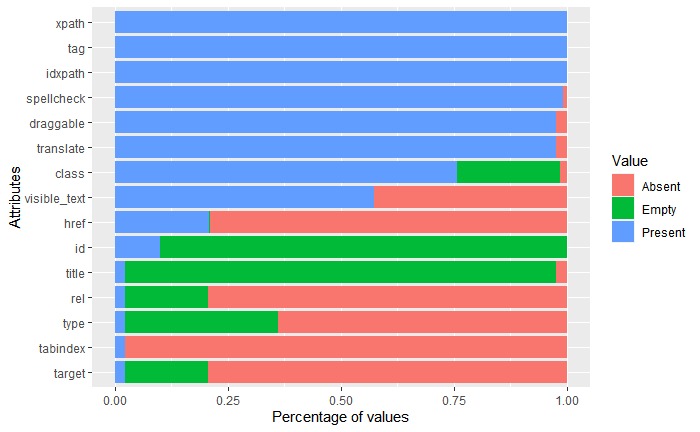}
    \caption{Distribution of absent, empty and valued attributes in the selected web pages}
    \label{fig:valid_values_new}
\end{figure}

\begin{figure}
    \centering
    \includegraphics[width=\columnwidth]{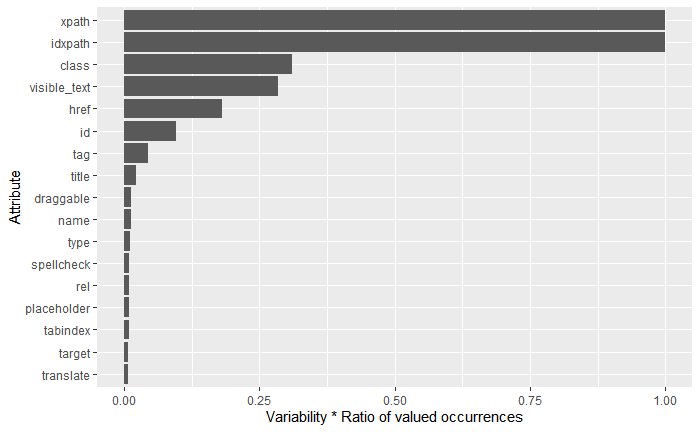}
    \caption{Top attributes for weighted variability in the selected web pages}
    \label{fig:variability_values_new}
\end{figure}

To analyze and justify the selection of attributes for the multi-locator-based algorithms, we computed statistics on the distribution of the attributes on all the included web applications.
We computed statistics for all the 170 standard HTML attributes listed by W3C\footnote{https://www.w3.org/wiki/Html/Attributes/\_Global}. 

In the considered web applications, we utilized only 35 attributes (i.e., they were assigned at least a non-null value).
In Figure \ref{fig:valid_values_new} we report the percentage of present, empty ("") or absent (\emph{null}) values for each attribute.
For the sake of readability, we only report the attributes for which the percentage of times a value is present is above the median for all attributes (0.7\%).
We distinguish between \emph{null} and \emph{empty} values since for specific attributes (e.g., the textual content of a web element, identified by the attribute \emph{visible\_text}), the absence of a value is itself a piece of information that we can use to locate a web element.

We compute the variability of each attribute as the ratio between the number of different values present in the set of considered web elements for a given attribute and the number of times the attribute is valued (i.e., other than null or empty).
To assign a weight to the variability with the relevance of the web elements in identifying web elements, we multiply it with the percentage of valued occurrences measured in the previous step.
In Figure \ref{fig:variability_values_new} we report the top attributes for variability in the considered web pages.
This result validates the original selection of attributes performed in the original Similo paper since only the XPath, ID Xpath, class, text, href, id, and tag attributes were those with the highest variability. 
 
It is worth noting that the XPath and ID XPath attributes have a value equal to 1 for the variability multiplied by the valued occurrence ratio, meaning that they are the only two attributes valued for every web element and for which no pair of web elements can have equal values.

\begin{center}\fbox{
\begin{minipage}[t]{0.96\linewidth}
\textbf{Answer to RQ1}: The analysis of Non-Null values and variability for attributes in web pages identify XPath, ID XPath, class, visible text, href, id, and tags as the attributes that are most likely to change when present with non-null values in elements of DOM models. The selection of attributes for the Similo algorithm is therefore confirmed as optimal for web elements multi-locators.
\end{minipage}
}
\end{center}

\begin{table}
    \centering
    \small
        \caption{Mean (SD) values of Precision, Recall and Accuracy over the five test sets for Similo at varying thresholds}
        \label{tab:similores}
    \begin{tabular}{lrrr}
    \toprule
                Threshold & Precision & Recall & Accuracy\\
    \midrule
    0.20 & 0.688 (0.024) & 0.965 (0.056) & 0.766 (0.036)\\
    \textbf{0.28} & 0.796 (0.029) & 0.898 (0.077) & 0.823 (0.045)\\
    0.40 & 0.892 (0.034) & 0.615 (0.121) & 0.770 (0.066)\\
    0.60 & 1.000 (0.006) & 0.265 (0.105) & 0.632 (0.052)\\
    0.80 & 1.000 (0.000) & 0.008 (0.012) & 0.503 (0.006)\\
     \bottomrule 
    \end{tabular}
    \label{tab:table1}
\end{table}

\subsection{RQ2 - Similo Performance}

Having finalized the selection of attributes to consider for comparing target and candidate web elements, we evaluated the original Similo  over the set of 1163 matches as defined in section C.5.

We performed 5-fold cross validation and optimized the threshold value, used to detect a match, on a training set and then evaluated at that threshold on the hold-out test set. The optimal threshold values were very stable over all folds; for simplicity we thus report results for the full set of web elements. 

Table \ref{tab:table1} reports the mean and standard deviation of precision (P), recall (R), and accuracy (A) at varying thresholds. The optimal threshold, selected to maximize accuracy on the test sets, is shown in bold.

With minimal variations between different folds, the optimal threshold for the algorithm is a rather low value of 0.28. Raising the threshold of the algorithm, in fact, lowers the number of comparisons that are signalled as matches, which reduces the number of false positives. This low value for the threshold may suggest that few attributes with stable values (over the selected set of 14) are sufficient to identify a match between the original and the new web element. Raising the threshold over 0.40 guarantees a high precision for the algorithm (over 90\%) at the expense of a rapidly increasing number of false negatives (i.e., number of candidates that are not matched while being correspondent to the target).


\begin{center}\fbox{
\begin{minipage}[t]{0.96\linewidth}
\textbf{Answer to RQ2}: applied on the test dataset, the original version of the Similo algorithm is capable of providing a mean 82.3\% accuracy when a match threshold of 0.28 is used.
\end{minipage}
}
\end{center}

\begin{table}
    \centering
    \small
        \caption{Mean (SD) values of Precision, Recall and Accuracy over the five test sets for VON Similo at varying thresholds}
        \label{tab:similoeqres}
    \begin{tabular}{lrrr}
    \toprule
                Threshold & Precision & Recall & Accuracy\\
    \midrule
    0.20 & 0.680 (0.011) & 0.991 (0.008) & 0.760 (0.011)\\
    \textbf{0.40} & 0.968 (0.007) & 0.922 (0.110) & 0.941 (0.052)\\
    0.60 &  1.000 (0.004) & 0.475 (0.210) & 0.738 (0.105)\\
    0.80 &  1.000 (0.000) & 0.030 (0.042) & 0.515 (0.021)\\
     \bottomrule 
    \end{tabular}
    \label{tab:table2}
\end{table}

\subsection{RQ3 - VON Similo performance}

In Table \ref{tab:table2}, we report the results for VON Similo, in the same format as for Similo above. Also here, the optimal threshold value was stable over the five folds.

By comparing the results in the two tables, it is clear that the optimal threshold is higher for VON Similo. This is likely since the comparison of multiple attribute values --- instead of the one-to-one comparison of the original Similo algorithm --- increases the likelihood that a candidate web element is deemed as matching with the target one. In other words, the increase of the size of the candidate space, as described in section III, requires a larger threshold or the number of false positives increases.

Even in this case the number of true positives found decays rapidly with increasing threshold, however we can observe a precision of 100\% at a threshold of 0.60, supported by a 47.5\% recall. We note that practitioners can choose to select a different threshold level that favors either precision or recall depending on the risks and costs they judge a false positive or a false negative to have. However, it is clear that VON Similo can achieve higher accuracy scores than the original Similo algorithm.

\begin{center}\fbox{
\begin{minipage}[t]{0.96\linewidth}
\textbf{Answer to RQ3:} applied on the test dataset, the Visually Overlapping Nodes-based version of the Similo algorithm (VON Similo) is capable of providing 94.1\% accuracy when a match threshold of 0.40 is used.
\end{minipage}
}
\end{center}

\begin{figure}
    \centering
    \includegraphics[width=\columnwidth]{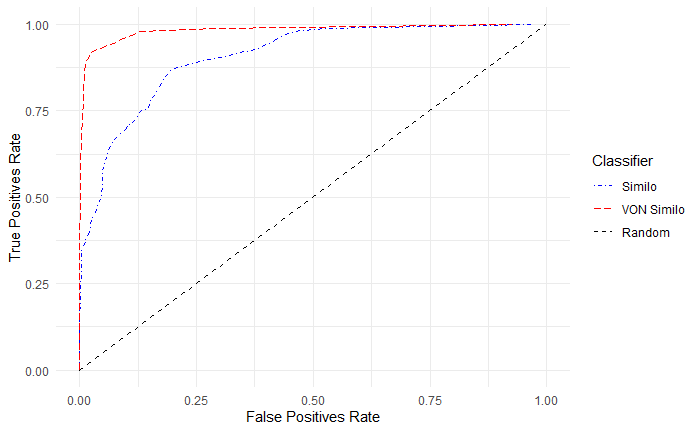}
    \caption{ROC comparison for Similo, VON Similo, and the baseline corresponding to a random classifier}
    \label{fig:rocs}
\end{figure}

\subsection{RQ4 - Comparison between Similo and VON Similo}

In Figure \ref{fig:rocs} we report the comparison of the two ROC (receiver operating characteristic) curves drawn for Similo (blue) and VON Similo (red). The ROC curves show the trade-off between the true positive rate (i.e. sensitivity) and the false positive rate (i.e. specificity) of an algorithm, plotted at varying thresholds. For simplicity, we report the ROC curves computed over the whole set of matches without applying the 5-fold split on the data set.

We also report in the graph, as a baseline, a random classifier, which is expected to provide points along the diagonal (i.e., number of true positives equal to the number of false positives). An ideal classifier would instead provide a single point where TPR = 1.00 and FPR = 0.00. In a ROC curve, the closer the curve is to the main diagonal of the ROC space, the less accurate the test.

For our purposes, there is no benefit in obtaining a Precision-Recall curve over a ROC curve since, by construction, the data are equally distributed between positives and negatives.

By visual comparison, it is evident how both the algorithms, Similo and VON Similo, provide ROC curves that are significantly distant from the main diagonal (corresponding to a Random classifier). At the same time, it is visually evident that the VON Similo algorithm is overall better than that of Similo. This advantage in using  VON Similo can be quantified by computing the area below the ROC Curve (Area Under Curve, or AUC).

\begin{table}
    \centering
    \small
        \caption{Comparison of the mean (std deviation) of precision, recall and accuracy of Similo vs. VON Similo, per subject application}
        \label{tab:similoeqres}
    \begin{tabular}{lrrr}
    \toprule
                Approach & Precision & Recall & Accuracy\\
    \midrule
    Similo & 0.799 (0.130) & 0.772 (0.243) & 0.783 (0.157)\\
    VON Similo & 0.965 (0.061) & 0.790 (0.915) & 0.879 (0.153)\\
     \bottomrule 
    \end{tabular}
    \label{tab:table3}
\end{table}

As a final comparison, in Table \ref{tab:table3}, we report a comparison of the performance of the two approaches in terms of average accuracy, precision and recall over the set of 33 web pages used in the experiment. From the comparison we notice that VON Similo has globally better values for all the measures on the set of apps, with 88\% accuracy against 78.3\% accuracy for Similo. We also observe rather high std. deviation values for the computed measures, caused by a variabile number of matches and performance obtained with the tool on individual applications.

One can also perform a Wilcoxon (paired) signed rank test~\cite{rey2011wilcoxon} of the accuracy values per app which supports (p-value of 0.0001614) the hypothesis that VON Similo (mean accuracy, over the apps, of 88.0\%)  performs differently than Similo (78.3\%). Since the argument has been made that Bayesian statistical analysis can have benefits~\cite{furia2019bayesian} we also compared the two approaches with a Bayesian signed rank test~\cite{benavoli2017time} which gave a probability of $99.9\%$ that VON Similo has a higher accuracy than Similo\footnote{There were clear differences also for precision and recall. We used Python 3.10 and the library baycomp for the Bayesian SignedRankTest and (base) R 4.2 for the Wilcoxon test.}.

\begin{center}\fbox{
\begin{minipage}[t]{0.96\linewidth}
\textbf{Answer to RQ4:} As demonstrated by the comparison of the respective ROC-curves and the statistical tests, VON Similo (AUC = 0.91, mean accuracy = 0.94, mean accuracy per app = 88.0) performs \textbf{significantly better} than Similo (AUC = 0.88, mean accuracy = 0.82, mean accuracy per app = 78.3).
\end{minipage}
}
\end{center}


\section{Discussion} \label{disc}
This study has shown that multi-locator-based approaches, using web element attributes, can provide robust web element localization, in the real world, for evolving web applications.
When coupled with visually overlapping node based (VON-based) locators, the method is robust with a 94 percent success rate in finding the correct web element.


This result is significant, especially when put into context of how tests are maintained in industry.
The study results are acquired from a random sample of web applications with both larger and smaller amounts of change between versions, controlled by the time in the past a web application was sampled (from 6 months to 5 years).
In a real scenario, in industry, the probability of only more minor changes between versions is theoretically higher than the scenario presented here since test suites would be run daily or at least weekly.
This leads us to the logical conclusion that; in an authentic setting, the average accuracy of the proposed approach would be higher than 94 percent.
We base this discussion on the attribute-based locator technology, which precision is affected by change.
For example, if only one attribute is changed, the accuracy will be higher than if two or more attributes have changed.
For shorter iterations between test runs, the number of changes to the application will be less and, as such, the solution accuracy increases.
However, although logically sound, future work is required to verify this conclusion, e.g., through empirical and industrial research.

Regardless, given the result of the study, an interesting question becomes what industry is willing to accept in terms of false test results, since, as stated in Section \ref{intro}, robust web element localization is one of the core challenges with automated GUI testing and perceived a root-cause to its lack of general use in practice.
Hence, although the presented results are significant, compared to, for instance, state of the art research by Leotta et al., which reported a success rate of 33-93 percent (93 percent being a theoretical best case limit)~\cite{leotta2015using}, a philosophical question remains if 94 percent success rate is good enough?
A 94 percent success rate still implies that six web elements will not be found every 100 test actions, or one faulty test behavior every 17 steps of script code run on modified parts of the SUT.
For a test suite ordering in the hundreds of test cases, 94 percent success rate would still constitute a significant amount of root-cause analysis and test script maintenance.
However, once more, the selected time elapsed between tests of the web application versions were in the experiment abnormal when compared to the time between executions of a GUI-based regression test suite in industrial practice.

However, as a speculative counterpoint, since any accuracy below 100 percent seems to be an issue for larger test suites, the question arises if 100 percent robustness is even achievable.
Keeping in mind that web element localization implies searching for a target web element in a context where there is often not an exact match, only a partial match, due to changes in the attributes, appearance, or behavior of the sought web element.
Under these circumstances, we also see humans having less than a 100 percent success rate, so the question is how far an automated approach can reach.

Additionally, the attribute-based approach shows that with impartial data, we can still identify web elements with high probability.
When one of these web elements is identified, the attributes that are no longer valid can be automatically repaired---Repair, in this case, implies updating attributes that are no longer valid.
Thus, mitigating much of the maintenance costs associated with test scripts.
Consequently, keeping the delta between test script versions and the SUT low implies a larger probability of web element identification success.
Thus, once more speaking towards the potential of the approach's ability to reach higher than 94 percent success rate in a real scenario in industrial practice.

Conceptually, VON-based locators can work for any DOM-based multi-locator approach, i.e., on other platforms not used in this paper, such as Android or desktop.
However, VON-based locators only work with multi-locator approaches and would not work for single-locator (e.g., Selenium-based) scripts due to a lack of information.
Consider a Selenium action based on an XPath locator, but said XPath is no longer available in a new version of the target web application.
All candidate information can be acquired from the new site, but if additional information is not available for the target web element, there would not be a way to map the target to any of the candidates reliably.
This is a limitation of the approach since, a priori, the amount of information to reliably locate a single-locator web element in a different version of the web application is not known.
This is a subject of future work that entails ranking the attributes based on their relative power to be used as locators.

Furthermore, the analysis of the attributes used for this work shows that the 14 attributes that Similo used are the most frequently populated ones on modern web pages.
Thus, implying that these provide the most value for web element localization.
This result is significant, as previous works have presented lists of suitable attributes~\cite{Kirinuki2019COLORCL, choudhary2011water, 10.1145/2950290.2950294}, but not provided empirical evidence to support these claims.
The list of attributes provided in this work, although only contemporary, thereby provides insights into how to design more robust GUI testing tools for the foreseeable future.
Future work could also investigate if there are notable dependencies, or even domain aspects, to how these attributes are populated and how, for further improvements into web element localization and automated repair.
Such analysis should also look at the relative power of different attributes.
Looking at the list of attributes in Figures \ref{fig:valid_values_new} and \ref{fig:variability_values_new} we note some interesting observations.
ID is an often mentioned attribute for web element localization~\cite{aldalur2020abla,leotta2015using,stocco2018visual}.
However, as shown by the results, ID is seldom used in practice, implying that reliance on this attribute would have detrimental effects on web element localization.
Hence, from a practical perspective, the attributes need to be weighed in terms of power versus availability.
One or several commonly available attributes, e.g., XPath or tag, have a higher probability of being useful than a perceived more powerful attribute, which is seldom available, such as ID.
Further research is, however, required to evaluate this relative power between attributes.

Another finding concerns the contents of the attributes of the web elements.
In Similo, locator parameters are compared with various comparators such as equals, Levenshtein distance, and integer distance.
These are utilized in this research based on expert judgment, but other comparators, e.g., euclidean distance, could also be used.
In future work, a valuable analysis would investigate the types and variability of attribute data to assign the most suitable comparators.
For instance, XPath strings are unique to each candidate and are represented as Strings of longer length.
This paper uses Levenshtein distance for their comparison, but other approaches would likely work equally well or even better.
This mapping between key web element attributes, the characteristics of the attribute data, and comparators could theoretically be done using machine learning, e.g., using learning to rank~\cite{xia2008listwise}.
Similarly, we can also optimize the weighting of each comparator-attribute tuple.
However, this is once more a subject of future research.

\subsection{Threats to validity} \label{threats}
In this section we discuss the threats to the study and its results pertaining to the internal, external and construct validity of the research.

\textbf{Generalizability:} The paper provides two contributions; (1) analysis of web element attributes used in industrial practice and (2) VON-based locators.
For (1), we perceive the results to be generally valid, as they are gathered from commonly used web pages from larger companies that should be perceived as adopters of best state-of-practice approaches.
However, for (2), the results are delimited to web element localization with multi-locators as discussed in Section \ref{disc}. 
This diminishes the result's industrial applicability since most industrial frameworks, like Selenium, utilize a single-locator approach.
As such, adopting VON-based multi-locators would also require a technological shift in approaches or tooling for developing such test cases.
Although scripts that use multi-locators can be developed manually, we perceive recording such tests as beneficial from an efficiency perspective.

\textbf{Subject selection:} We took the subjects for this analysis from the Alexa.com web application (as of writing, the site has been discontinued), which reduces bias in the sampling as the web applications on the list are outside the researchers' control.
However, we chose only a subset of the list, i.e., the top 40 pages (33 used after excluding, for instance,  broken and duplicate sites).
This set is still perceived as significant when compared to other studies in this area of research.
However, it is unknown how representative the result is to any general web application on the internet.
Analysis of the tags used in the web elements of these web pages does, however, provide us with confidence that all types of web elements that can, theoretically, be encountered have been identified.
We stress, however, that this research only focused on web element identification, and therefore detrimental effects that may arise during test execution were not covered.
An example of such an effect is the acquisition of web element attribute data during a state transition.
A lack of synchronization between the test script and AUT could result in missing attributes if pre-analysis of web element information begins before the AUT is fully loaded.

\textbf{Comprehensiveness:} VON Similo was designed with flexibility in mind to optimize its potential use in practice.
A drawback of this design is that the comparators and attributes used in this study are a subset of possible attributes and comparators, as discussed in Section \ref{disc}.
While not reducing the validity of the results presented in this work, this implies that we could achieve better results with other constellations of parameters.
We defend this unorthodox statement by the presented results that indicate a considerable increase in precision and recall compared to state of the art.
As such, although there is no threat to the validity of the results of the current implementation, given the set of analyzed web pages, we cannot state if this is the best implementation of the approach.
Further research is thereby required with other constellations and contexts to optimize the approach.

\section{Conclusions and Future Work} \label{conc}

In this paper, we proposed a novel location algorithm for web elements in web applications, i.e., Visually Overlapping Nodes (VON). We applied this approach by extending an existing multi-locator approach (Similo), thereby obtaining the VON Similo approach.

To investigate the accuracy of the algorithm, we performed an empirical investigation by manually identifying a set of matches between mutated widgets in different releases of the same application and then measuring the effectiveness of the original and extended approaches to determine correct matches.
This analysis led us to measure a +9.9\% increase in accuracy for VON Similo with respect to the original approach.
It is worth underlining that the original Similo was already proven as better performing than state-of-the-art multi-locator algorithms (e.g., Leotta's ROBULA+ \cite{leotta2016robula+}).

This work first assessed the possible benefits introduced by the concept of Visually Overlapping Nodes when performing GUI testing of web applications.
Therefore, fine-tuning the approach can be obtained by empirically finding the optimal values for several fixed coefficients and variables employed by the algorithm.

The algorithms evaluated and compared in this study used only static weights defined in the original Similo algorithm and were compatible with state-of-the-art tools like COLOR\cite{Kirinuki2019COLORCL}. Selecting a different set of weights for the attributes used in the comparison can lead to significantly different results in terms of precision and accuracy. The same reasoning applies to selecting the attributes involved in the score computation.

In our immediate future work, we plan to employ machine learning-based approaches to identify the optimal set of attributes, and related weights, to maximize the accuracy of Similo.
As a further improvement of the algorithm, it is possible to consider returning as output not only a single matching web element but a ranked list of the most matching candidates.
This evolution of the algorithm can also benefit from the application of machine-learning-ranking (MLR or Learning to Rank) techniques.

\bibliography{bibfile}



\end{document}